\documentclass[fleqn,usenatbib]{mnras}

\usepackage{newtxtext,newtxmath}

\usepackage[T1]{fontenc}

\DeclareRobustCommand{\VAN}[3]{#2}
\let\VANthebibliography\thebibliography
\def\thebibliography{\DeclareRobustCommand{\VAN}[3]{##3}\VANthebibliography}

\usepackage{graphicx}	
\usepackage{amsmath}	
\usepackage{xcolor}

\newcommand{\Sighi}{\Sigma_{\mathrm{HI}}}
\newcommand{\HI}{\ensuremath{\mathrm{H}\scriptstyle\mathrm{I}}}
\newcommand{\sighi}{\sigma_{\mathrm{HI}}}

\newcommand{\vproj}{v_{\mathrm{proj}}}
\newcommand{\hiflux}{\Psi_{\mathrm{tot}}}
\newcommand{\lcdm}{$\Lambda$CDM}

\title[Testing \HI{} flux profiles]{On the consistency of rotation curves and spatially integrated \HI{} flux profiles}

\author[Yasin \& Desmond]
{Tariq Yasin$^{1}$\thanks{\href{mailto:tariq.yasin@physics.ox.ac.uk}{tariq.yasin@physics.ox.ac.uk}}
and Harry Desmond$^{2}$
\\
$^1$Astrophysics, University of Oxford, Denys Wilkinson Building, Keble Road, Oxford, OX1 3RH, UK\\
$^2$Institute of Cosmology \& Gravitation, University of Portsmouth, Dennis Sciama Building, Portsmouth, PO1 3FX, UK\\
}

\date{Accepted XXX. Received YYY; in original form ZZZ}

\pubyear{2015}

\begin{document}
\label{firstpage}
\pagerange{\pageref{firstpage}--\pageref{lastpage}}
\maketitle

\begin{abstract}
Resolved rotation curves (RCs) are the gold-standard measurements for inferring dark matter distributions in $\Lambda$CDM and testing alternative theories of dynamics in galaxies. However they are expensive to obtain, making them prohibitive for large galaxy samples and at higher redshift. Spatially integrated \HI{} flux profiles are more accessible and present the information in a different form, but---except in a highly compressed form, as linewidths---have not so far been compared in detail with RCs or employed for dynamical inferences. Here we study the consistency of RCs and \HI{} surface density profiles from SPARC with spatially integrated \HI{} flux profiles from ALFALFA, by combining the resolved properties in a forward model for the flux profile. We define a new metric for asymmetry in the flux profiles, enabling us to cleanly identify those unsuitable for our axisymmetric method. Among all SPARC galaxies the agreement between RCs and flux profiles is satisfactory within the limitations of the data---with most galaxies having an uncertainty-normalised mean squared error (MSE) below 10---while no galaxy deemed symmetric has a MSE above 1.2. Most cases of good agreement prefer an \HI{} gas dispersion $\sighi$ of \textasciitilde13 km/s, consistent with resolved studies of gas dispersion from the literature.
These results open the door for spatially integrated \HI{} flux profiles to be used as proxies for spatially resolved dynamics, including a robust appraisal of the degree of asymmetry.
\end{abstract}




\section{Introduction}

Neutral hydrogen gas (\HI{}) is a crucial probe of the kinematics of galaxies. It is the most extended readily-detectable component of baryonic matter in late-type galaxies, reaching into the outer regions where dark matter is thought to dominate the dynamics. Further, its dynamical coldness allows for relatively simple analyses compared to stars or warmer gas. Resolved studies of \HI{} (e.g.~\citealt{walterTHINGSHINearby2008,ponomarevaDetailedKinematicsTullyFisher2016,lelliSPARCMASSMODELS2016}), typically measure \emph{rotation curves} (RCs), the azimuthally-averaged
rotation velocity of the gas as a function of radius from the centre of the galaxy. RCs are powerful probes of the mass distribution around galaxies, and have been used extensively to measure dynamical scaling relationships \citep[e.g.][]{mcgaughRadialAccelerationRelation2016,ponomarevaLightBaryonicMass2018,lelliBaryonicTullyFisher2019}, assess the consistency of galaxy dynamics with \lcdm{} expectations \citep[e.g][]{desmondTullyFisherMasssizeRelations2015,katzTestingFeedbackmodifiedDark2017,liComprehensiveCatalogDark2020,postiDynamicalEvidenceMorphologydependent2021,mancerapinaImpactGasDisc2022,yasinInferringDarkMatter2023,stiskalekFundamentalityRadialAcceleration2023}, and test alternative dark matter models \citep[e.g.][]{2022arXiv220710638A,misteleGalacticMasstolightRatios2022,khelashviliDarkMatterProfiles2023} and theories of gravity \citep[e.g.][]{milgromModificationNewtonianDynamics1983,burrageRadialAccelerationRelation2017,naikConstraintsChameleonGravity2019,desmondTensionRadialAcceleration2024}.

However, RCs are observationally expensive to measure, requiring both adequate spatial resolution
and long integration times to obtain enough signal in each radial bin. Although ongoing and forthcoming surveys \citep[e.g.][]{maddoxMIGHTEEHIHIEmission2021,vancappellenApertifPhasedArray2022,koribalskiWALLABYSKAPathfinder2020}, utilising instruments such as the Square Kilometre Array (SKA), will dramatically improve the instrumental resolution and sensitivity available to RC studies, they will remain either impossible or impractical to obtain in some of the regimes of greatest scientific interest. In terms of reaching higher redshift, \HI{} RCs will still only be obtainable out to z\textasciitilde0.5 \citep{blythExploringNeutralHydrogen2015} even with the SKA. For probing the faint end of galaxy formation using nearby galaxies, the angular resolution of future instruments will still only be able to resolve the closest of the lowest mass dwarf galaxies, and even then at great observational expense \citep{maddoxMIGHTEEHIHIEmission2021}.

In contrast to RCs, spatially integrated \HI{} flux profiles are a cheap and efficient way to probe the kinematics of cosmological volumes of galaxies. Typically these are summarised by statistics such as the linewidth, the width of the profile at some fraction of the mean or peak flux. Linewidths have been used to study the Tully--Fisher relation \citep[e.g.][]{lelliBaryonicTullyFisher2019, ponomarevaMIGHTEEHBaryonicTully2021,ballGeneralistAutomatedALFALFA2023, gogateBUDHIESBaryonic2023}, constrain dark matter halo properties \citep{yasinInformationHaloProperties2023} and assess the consistency between \lcdm{} and dwarf kinematics in cosmological simulations \citep{sardoneClosingGapObserved2023} and semi-analytical models (SAMs;~\citealt{brooksNorthsouthAsymmetryALFALFA2023}).

HI flux profiles exhibit a diverse range of shapes, typically simplified into two main categories, although in reality there is a continuum between them \citep{obreschkowSimulationCosmicEvolution2009a,stewartSimpleModelGlobal2014}. The classic ``double-horned'' profile occurs due to flux buildup at the wavelengths corresponding to the velocity of the flat part of the RC on the approaching and receding sides of the galaxy. Single-peaked, closer to Gaussian-shaped profiles tend to occur in lower mass galaxies where the \HI{} does not probe into the flat part of the RC, or when the horns are smeared out by either gas dispersion or instrument noise (which can become comparable to the projected rotational velocity in low mass galaxies), or galaxies with close to face-on orientations. In total, the observed flux profile clearly depends on both the resolved dynamics of the system, as well as the spatial distribution of \HI{} and galaxy orientation. 

Naturally, some information on the dynamics is lost in the data compression from the full flux profile to the linewidth. Further, there is debate around which definition of the linewidth is most informative and reliable, as well as the best way to calculate it from the raw spectrum whilst accounting for observational effects such as noise and instrumental resolution \citep[e.g.][]{westmeierBusyFunctionNew2014,yuDeterminationRotationVelocity2020,pengParameterizedAsymmetricNeutral2023,ballGeneralistAutomatedALFALFA2023,sardoneClosingGapObserved2023}. As the linewidth is typically defined as the width measured at some percentile of the mean or maximum of the flux, it depends on the shape of the flux profile. Cuts to select Tully--Fisher samples are made based on inspection of flux profiles \citep[]{tullyCosmicflows2Data2013}, and corrections are made to try and link the linewidth to twice the maximum rotation velocity \citep{verheijenUrsaMajorCluster2001}. Details of the flux profile are also important for pipelines such as \texttt{Bayestack} \citep{panMeasuringBaryonicTullyFisher2021}, which make assumptions on the shape of the profile in order to recover the \HI{} signal from sub-noise threshold measurements.

More recently the \emph{shape} of the flux profile has been studied theoretically in cosmological simulations \citep{el-badryGasKinematicsFIRE2018a, glowackiASymbaGlobalProfile2022} and semi-analytical models (SAMs) \citep{2021arXiv210504570P}. \citeauthor{2021arXiv210504570P} and \citeauthor{el-badryGasKinematicsFIRE2018a} highlighted the linewidth as a potential probe of the kinematics with more information than the linewidth. \cite{glowackiASymbaGlobalProfile2022} show that the asymmetry of flux profiles could also be used to test galaxy formation processes such as mergers. Observationally, the asymmetry of flux profiles has been measured and correlated with galaxy properties in large surveys \citep[e.g.][]{yuStatisticalAnalysisProfile2022}. \citet{papastergisAccurateMeasurementBaryonic2016} found that applying cuts based on flux profile asymmetry changed the properties of the baryonic Tully--Fisher relation measured using \HI{} linewidths.

Clearly, understanding the relationship between the integrated \HI{} flux profile and resolved properties of galaxies is important for understanding the kinematic information derivable from integrated profiles, to formulate tests of dark matter and/or galaxy formation which depend on it, and to understand derived products better. Whilst the relationship between resolved dynamics and integrated flux profiles has been studied theoretically in simulations and SAMs, it has not yet been studied \emph{empirically}. To address this gap, in this work we study the consistency between resolved rotation curves and \HI{} surface density profiles from the SPARC database \citep{lelliSPARCMASSMODELS2016}, and integrated \HI{} flux profiles as observed by the ALFALFA survey \citep{haynesAreciboLegacyFast2018} in a manner independent of the galaxies' dark matter mass distributions. This will test to what extent the flux profile can be used as a proxy for the resolved RC, and highlight potential biases in observed flux profiles that may be relevant for ongoing and future studies.

In Section \ref{sec:data} we describe the data used in this study. In Section \ref{sec:methods} we describe the model used to relate the \HI{} flux profile to the resolved properties of the galaxy, and the statistics we use to assess their consistency. In Section \ref{sec:results} we present the results of our comparison. In Section \ref{sec:discussion} we discuss the implications of our results, and in Section \ref{sec:conclusions} we present our conclusions. All logarithms are base-10.

\section{Data}\label{sec:data}

\subsection{SPARC}\label{sec:SPARC}

For our resolved data we use the SPARC database\footnote{\url{http://astroweb.cwru.edu/SPARC/}}, which contains \HI{} rotation curves for 175 late-type galaxies \citep{lelliSPARCMASSMODELS2016}. We also use their azimuthally-averaged \HI{} surface density profiles (F. Lelli, priv. comm.), which were available for 169 galaxies. The data in the SPARC database is compiled from around three decades of literature. The RCs are measured using 21-cm hydrogen emission, with the addition of H$\alpha$ to probe the inner regions of some galaxies.  The SPARC compilation spans a broad spectrum of galactic properties, with luminosities ranging from \(10^7 \: L_{\odot}\) to \(10^{12} \: L_{\odot}\), surface brightness from approximately 5 to \(5000\: L_{\odot} \,\mathrm{pc}^{-2}\), and \HI{} masses between \(10^7\) and \(10^{10.6} \: M_{\odot}\), and a broad range of morphologies. The database provides a quality flag (1:high, 2:medium, 3:low). We only use quality 1 and 2 galaxies in our analysis.

\subsection{ALFALFA}\label{sec:ALFALFA}

We take integrated flux profiles from the ALFALFA\footnote{\url{http://egg.astro.cornell.edu/alfalfa/data/index.php}} \citep{haynesAreciboLegacyFast2018} data set, which was obtained by a blind survey with the Arecibo telescope across 7000 \(\text{deg}^2\) of the Northern sky, with detections extending to a redshift of $\sim$0.06. The final \(\alpha.100\) catalogue contains approximately 31,500 extragalactic sources, with a \(\text{H\textsc{i}}\) mass range from \(10^6\) to \(10^{11}\) \(\textrm{M}_{\odot}\). The reduced data is provided as a baseline-subtracted spectrum for each galaxy with flux given in velocity bins of $\sim$5 km/s.

We perform a crossmatch with SPARC optical positions (obtained from the NASA/IPAC Extragalactic Database\footnote{\url{https://ned.ipac.caltech.edu}}), requiring matches to be within 4 arcseconds. We find 40 galaxies in common between the two datasets, and perform a basic check of the matching by comparing the SPARC and ALFALFA \HI{} masses and linewidths. We drop NGC4214 from the sample as its \HI{} mass reported by both the SPARC \citep{lelliDynamicsStarburstingDwarf2014a} and THINGS \citep{deblokHighResolutionRotationCurves2008} surveys are around 10 times the ALFALFA value (when assuming the same distances), and also disagree with the ALFALFA linewidth.

The fiducial ALFALFA pipeline is optimised to recover accurate flux profiles for unresolved sources. However many of the SPARC galaxies are nearby, and hence have angular extent significantly greater than the ALFALFA beam width. This leads to flux in the galaxy outskirts being missed by the original pipeline, which can result in one or both horns of the flux profile being artificially absent (i.e. without indicating a physical asymmetry).

\cite{hoffmanTotalALFALFANeutral2019} produced an alternative ALFALFA analysis pipeline to allow the extraction of precise fluxes for galaxies larger than a few arcminutes on the sky.  We confirm by visual inspection that the horns are more prominent for many of the reanalysed galaxies, showing that the reanalysis has picked up flux from the outskirts that was missed by the original pipeline. The full spectrum data was not kept for all galaxies reanalysed by  \citeauthor{hoffmanTotalALFALFANeutral2019} (which mainly focused on the total flux). Therefore our final sample of 20 galaxies consists of 10 galaxies for which the \citeauthor{hoffmanTotalALFALFANeutral2019} reanalysed spectra were available (L. Hoffman, priv. comm.), and 10 galaxies which have too small an angular extent on the sky for the reanalysis pipeline to be necessary.

\section{Methods}\label{sec:methods}

\subsection{Calculating an empirical flux profile from resolved  data}\label{sec:model}

Both the \HI{} surface density as a function of radius, $\Sighi(r)$, and the projected rotation curve, $v_{\mathrm{proj}}(r)=v_{\mathrm{rot}}(r)\sin(i)$ (where $v_{\mathrm{rot}}$ is the intrinsic, inclination ($i$)-corrected circular velocity) are measured in resolved RC studies. Therefore a direct way to assess the consistency of SPARC RCs and ALFALFA flux profiles is to calculate an \emph{empirical} flux profile using the SPARC observed $\Sighi$ and $\vproj$, which can be compared to the ALFALFA data. We note that as $v_{\mathrm{proj}}$ is the relevant quantity for calculating the flux profile, no correction for $i$ is needed.

To calculate a model flux profile from the azimuthally-averaged RC and $\Sighi$, we follow \cite{obreschkowSIMULATIONCOSMICEVOLUTION2009} in modelling the \HI{} as a series of concentric infinitely thin rings of gas following circular orbits with velocity $v_{c}(r)$ and with surface density $\Sigma_{\HI}(r)$. The normalised flux per unit velocity emitted from a single ring of gas is
\begin{equation}
    \tilde{\psi}\left(v_{\lambda}, v_{\mathrm{proj}}\right)=\left\{\begin{array}{ll}
    \frac{1}{\pi \sqrt{v_{\mathrm{proj}}^{2}-v_{\mathrm{\lambda}}^{2}}} & \text { if }\left|v_{\mathrm{\lambda}}\right|<v_{\mathrm{proj}} \\
    0, & \text { otherwise }
    \end{array}\right.
    \end{equation}
where $v_{\lambda}$ is the velocity with respect to the barycentre of the galaxy (corresponding to an observed wavelength $\lambda$) and $\int \tilde{\psi} dv = 1$. We introduce a velocity dispersion to the gas to account for its random motion (which also smooths the singularity in the above equation at $v_{\mathrm{proj}}=v_{\lambda}$), so the normalised flux emitted from a single ring of gas is

\begin{equation}
    \psi\left(v_{\mathrm{\lambda}}, v_{\mathrm{proj}}\right)=\frac{\sigma_{\HI{}}^{-1}}{\sqrt{2 \pi}} \int^{+\infty}_{-\infty} d V \exp \left[\frac{\left(v_{\mathrm{\lambda}}-V\right)^{2}}{-2 \sigma_{\HI{}}^{2}}\right] \tilde{\psi}\left(V, v_{\mathrm{proj}}\right)
    \end{equation}
where $\sigma_{\HI{}}$ is the width of the Gaussian dispersion. The overall flux per unit velocity is then the integral of $\psi$ over the whole galaxy, weighted by the surface density

\begin{equation}\protect\hypertarget{eq:convolution}{}{
    \psi_{\mathrm{tot}}\left(v_{\lambda}\right)=\frac{2 \pi}{M_{\mathrm{HI}}} \int_{0}^{\infty} dr \: r \Sigma_{\mathrm{HI}}(r) \psi\left(v_{\lambda}, v_{\mathrm{proj}}(r)\right),
    }\label{eq:convolution}\end{equation}
where the normalisation gives $\int  \psi_{\mathrm{tot}} dv = 1$. The observed 21-cm flux profile can be calculated from this using the standard formula $\Psi_{\mathrm{tot}} = \frac{M_{\mathrm{HI}}}{D^2}  \frac{\mathrm{ Jy \,  kms^{-1} \, Mpc^2 M_{\odot}^{-1}}}{2.356 \times 10^5}$ \citep{haynesAreciboLegacyFast2018}, with $\Psi_{\mathrm{tot}}$ in Jy km/s, $D$ the assumed distance in Mpc and $M_{\mathrm{HI}}$ in $M_{\odot}$. Finally, to account for instrumental broadening we convolve the profile with a Gaussian kernel with full-width-half maximum equal to the 5 km/s instrumental resolution \citep{verheijenUrsaMajorCluster2001}.

\begin{table}
    \centering
    \caption{\label{tab:parameters}The free parameters in our model for the  \HI{} flux profile, which combines the SPARC RC and \HI{} surface density profile according to equation \ref{eq:convolution}. For both parameters we use a wide uniform prior, finding the posterior to be constrained to well within the prior range.}
    \begin{tabular}{ |c|c|c|c|c| }
      \hline
       & \textbf{Parameter} & \textbf{Units} & \textbf{Definition} & \textbf{Prior} \\

      \hline
        & $\sighi$ & km/s & \HI{} velocity dispersion & Uniform \\
        & $\hiflux$ & Jy km/s & Total \HI{} flux & Uniform \\
      \hline

    \end{tabular}
\end{table}

For some galaxies the measured $\Sighi$ extends beyond the measured $\vproj$. For cases where the galaxy has reached the flat part of its RC, we extrapolate $\vproj$ with a fixed value equal to $v_{\textrm{flat}}$, the mean velocity of the flat part of the RC (defined\footnote{The algorithm to identify the flat part is iterative, starting with the outermost data points, and adding additional points until the next point deviates in velocity by more than 5\% from the mean of the current points. A minimum of three included points is required for the RC to be considered as having a flat part.} in \citealt{lelliSMALLSCATTERBARYONIC2015} and tabulated in SPARC). For the rest of the galaxies we linearly extrapolate using the last three data points (we also test extrapolating with a fixed value equal to the last measured point, finding that it does not change our results). It is also expected that, to some degree, the radial extent of the gas probed by integrated observations will be greater than the maximum radius of resolved $\Sighi$ measurements, due to the finite sensitivity of the latter.
We do not attempt to remedy this by extrapolating $\Sighi(r)$ (which would be highly uncertain), but we investigate the impact of the differing sensitivities of the integrated and resolved observations, and the extrapolation of both the RC and $\Sighi$, in Section \ref{sec:results}.

\subsection{Inference}

We introduce two free parameters: $\sighi$, the \HI{} velocity dispersion and $\hiflux$, the total \HI{} flux, which are varied to minimise the difference between the predicted and observed flux profiles. For any given instrument there is a 10\% calibration uncertainty in sensitivity to 21-cm emission \citep[e.g.][]{haynesAreciboLegacyFast2018}, which dominates over the statistical uncertainty for galaxies in our sample (which have high signal-to-noise ratio). Therefore the calibration uncertainty on both ALFALFA and SPARC instruments is absorbed into $\hiflux$ in our model, which we give a wide flat prior. Observations of nearby spirals have estimated the distribution in the mean $\sighi$ across the galaxy of $12 \pm 3$ km/s \citep{mogotsiHICOVelocity2016}, which we initially adopt as a Gaussian prior on the $\sighi$ model parameter. However we find the data is constraining enough that the preferred $\sighi$ is almost identical using either a Gaussian or uniform prior, so adopt the latter. We give $\Psi_{\mathrm{tot}}$ a wide uniform prior between the value tabulated in ALFALFA $\pm 25\%$, which is found by simply integrating the flux profile. The parameters are summarised in Table \ref{tab:parameters}.

The likelihood of the ALFALFA flux profile for an assumed $\sighi$, $\hiflux$, $\Sighi$(r) and $\vproj$(r) is
\begin{equation}\label{eq:hi_likelihood}\protect{}{
\mathcal{L}(\mathcal{D}|\theta,\mathcal{M}) = {\displaystyle \prod_{j}^{n}} \frac{\exp\{
-(\Psi_{\text{j,obs}} - \Psi_{\text{pred}}(v_{\lambda,j},\vproj))^2 / (2 \sigma_\text{rms}^2) \}}
{\sqrt{2\pi}\sigma_\text{rms}} },
\end{equation}
where $\mathcal{D}$ is the data, $\theta$ the parameter vector and $\mathcal{M}$ the model. $\Psi_{\text{j,obs}}$ is the observed \HI{} flux in velocity bin $j$ (of $n$ total) and $\Psi_{\text{pred}}(v_{\lambda,j})$ the model prediction. The observational uncertainty $\sigma_\text{rms}$ is the $rms$ noise, calculated on regions free of signal and radio frequency interference around each flux profile.


To marginalise over the reported uncertainties on $\vproj$ (which we assume to be Gaussian and uncorrelated), at each step of the inference we draw 500 Monte Carlo samples of $v_{\mathrm{proj}}(r)$ (from the Gaussian distributions set by the errorbars), and take the product of their individual likelihoods to be the total likelihood (we check convergence with respect to number of samples). There is no information on the uncertainties on the individual points of $\Sighi$, so we do not include these in our analysis, but the statistical uncertainty is expected to be subdominant.

We use the \texttt{emcee} package \citep{foreman-mackeyEmceeMCMCHammer2013} to sample the posterior, using 20 walkers and a stretch move of $a=2$. We ensure the chain is well converged by iterating until the chain length is 50 times the autocorrelation length and checking the posteriors are stable with respect to further iteration.

\subsection{Assessing consistency}

We quantify agreement between model and data using the standardised mean squared error :

\begin{equation}\label{eq:residual}
    \mathrm{MSE} = \frac{1}{n} \sum^n_{j} {\frac{(\Psi_{\text{j,obs}} - \Psi_{\text{j,pred}})^2}{\sigma_\text{rms}^2}}.
\end{equation}
All models are able to fit the signal-free region surrounding the 21-cm line well, so agreement improves as more of it is included in the statistic. Therefore we define the signal region as the region around the line centre containing no points within 0.25 $\sigma_\text{rms}$ of zero, and only use this to calculate MSE (the actual fit still includes a large signal-free region around the profile).
Our procedure for determining this is the following.
Beginning at the line centre, we move outwards on each side independently, adding the next point to the signal region if it is not within 0.25 $\sigma_\text{rms}$ of zero. We stop (independently on each side) upon reaching the first point with flux greater than 0.25 $\sigma_\text{rms}$.

\subsection{Quantifying asymmetry}

We quantify asymmetry in the observed flux profile in order to identify galaxies for which our assumption of an axisymmetric disc may be too restrictive. To do this we fit the observed flux profiles with the ``generalised busy function'' of \citet{westmeierBusyFunctionNew2014}, which was designed to be able to fit observed flux profiles well. The function is

\begin{align}
    B(x) &= \frac{a}{4}
            \times (\mathrm{erf}[b_{1} \{ w + x - x_{\rm e} \} ] + 1) \nonumber \\
         &  \times (\mathrm{erf}[b_{2} \{ w - x + x_{\rm e} \} ] + 1)
            \times \left( c \, |x - x_{\rm p}|^{n} + 1 \right) \!,
    \label{busyfunction}
\end{align}
where $a,b_1,b_2,x_e,x_p,c,w,n$ are free parameters and $\mathrm{erf}$ is the error function. The function can be made symmetric by setting $b_1=b_2$ and $x_p=x_e$. Therefore we quantify asymmetry as the change in Bayesian Information Criteria (BIC, \citealt{schwarzEstimatingDimensionModel1978}) for each galaxy between using the full 8-parameter fit and the symmetric 6 parameter fit

\begin{equation}
    \Delta{\mathrm{BIC}_{\mathrm{asym}}} = \mathrm{BIC}_{\textrm{symmetric}} - \mathrm{BIC}_{\textrm{asymmetric}}.
\end{equation}\label{eq:bic}
A higher $\Delta{\mathrm{BIC}_{\mathrm{asym}}}$ corresponds to a more asymmetric profile, allowing one to choose a threshold corresponding to one's tolerance for asymmetry. Here we consider a galaxy to be asymmetric if it has $\Delta{\mathrm{BIC}_{\mathrm{asym}}} >  10$, corresponding to ``very strong'' evidence for asymmetry on the Jeffreys scale \citep{jeffreysTheoryProbability1998}.

\section{Results}\label{sec:results}

\begin{figure*}
    \includegraphics[height=0.16\textheight]{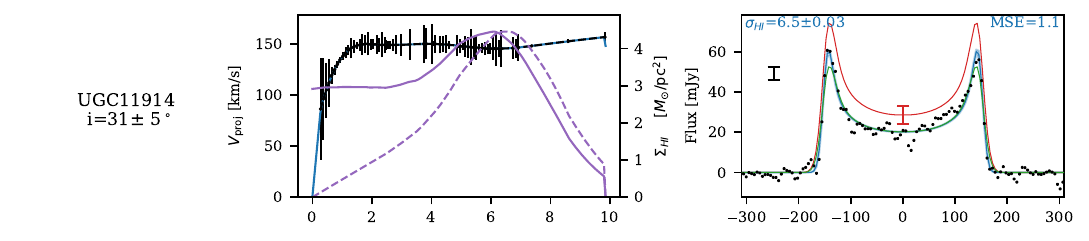}
    \includegraphics[height=0.16\textheight]{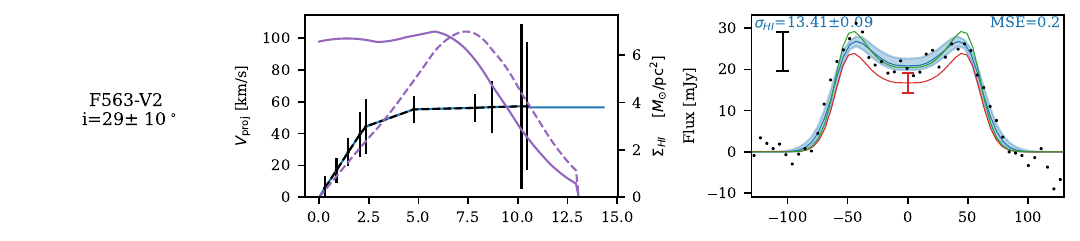}
    \includegraphics[height=0.16\textheight]{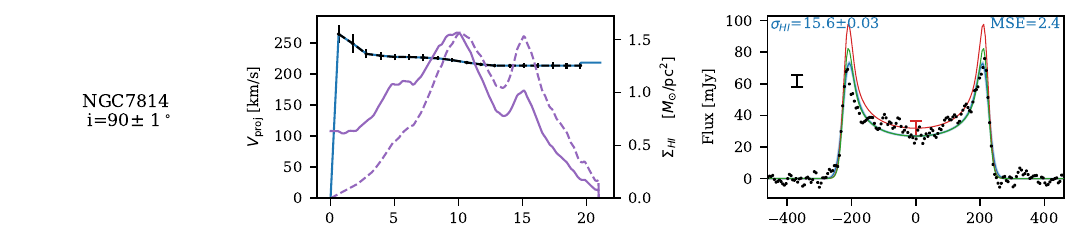}
    \includegraphics[height=0.16\textheight]{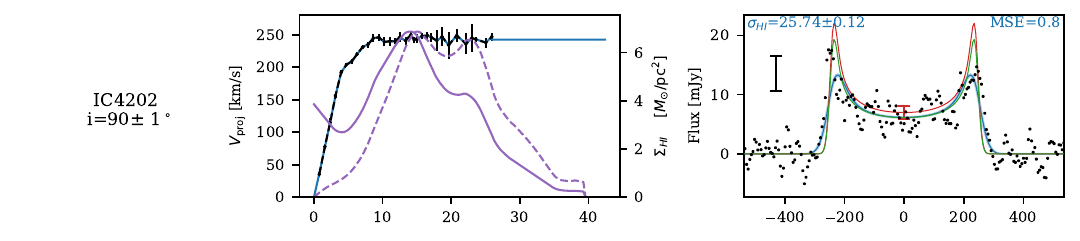}
    \includegraphics[height=0.16\textheight]{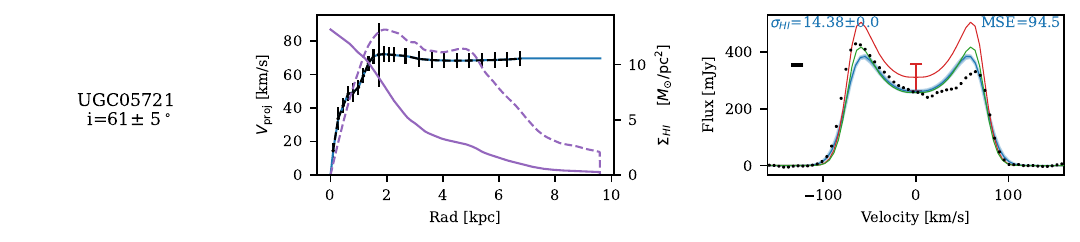}

    \caption{Results for 5 example galaxies. \\
    \emph{Left panels:} The SPARC data: the observed RC (black) (with its extrapolation in blue, see Section \ref{sec:model}), the \HI{} surface density $\Sighi$ (purple) and $r$$\Sighi$ (dashed purple). The latter is proportional to \HI{} flux per unit radius, and hence shows which region of the RC the flux profile is sensitive to. \\
    \emph{Right panels:} The best fit model flux profile (blue, with 1$\sigma$ uncertainty shaded) and the ALFALFA flux profile (black). The model is calculated by combining the SPARC RC and $\Sighi$ according to equation \ref{eq:convolution}. The preferred $\sighi$ and its uncertainty are labelled in blue, as well as the MSE goodness-of-fit statistic (equation \ref{eq:residual}). Green is the best fit model, but with $\sighi$ fixed to the literature expectation of 12$\pm$3 km/s (with dashed lines the uncertainties). Red shows the best fit model, but with $\hiflux$ fixed to that implied by SPARC; the errorbar corresponds to the 10\% calibration uncertainty. The black vertical errorbar is the uncertainty on the flux measurements.}
    \label{fig:empirical-fit}
\end{figure*}

\subsection{Overview}\label{sec:overview}
In Fig.\ref{fig:empirical-fit} we show the data and our fits to the flux profile for some example galaxies. In the left panel we show the SPARC observed RC (black) and the extrapolated RC used as input to our model (blue). In purple is the observed $\Sighi$, and the dashed purple is proportional to $r \Sighi$, the total \HI{} flux per unit radius, which shows which radii the integrated flux profile is most sensitive to. The integrated flux profile is not sensitive to the lowest radii, with the sensitivity typically highest at some point of the flat part of the RC before dropping off at large radius. For IC4202 the RC is extrapolated significantly as $v_{\mathrm{flat}}$, in order to extend it as far as the $\Sighi$ data.

In the right panel we show the best fit models to the flux profiles. Both $\hiflux$ and $\sighi$ are very well constrained for all galaxies, so we do not show their posteriors. However we label the favoured $\sighi$ on the plot with its uncertainty. Due to the tight parameter constraints, the uncertainty on the model flux profile (shown by the shaded blue bland) is dominated by the uncertainty on the SPARC RC. In red we show the best fit from the inference, but with the total \HI{} flux scaled to match that implied by the SPARC $M_{\mathrm{HI}}$, along with a $10\%$ uncertainty corresponding to the calibration error. This shows how much the flux profile is sensitive to the normalisation of the flux profile, and potential differences in the flux measured by SPARC and ALFALFA.

In green we show the flux profile when $\hiflux$ is fixed to the value found in the inference, but $\sighi$ is given a Gaussian distribution of  $12 \pm 3$ km/s corresponding to literature expectations. The uncertainty, shown by the dashed green lines, now corresponds to the range of $\sighi$ (in addition to the RC uncertainty), so is larger. We see from the bottom three panels that the effect of increasing $\sighi$ is to dampen the horns and broaden the wings of the flux profile.

For the top two galaxies (UGC11914 and F563-V2) the model flux profile is a good match to the data, and the preferred $\sighi$ is consistent with expectations. For NGC7814 the model is a fairly good match to the data, with a MSE of 2.4 and a reasonable $\sighi$. For IC4202 the model is a good match to the data, but a higher $\sighi$ is preferred than expected. UGC05721 has a strongly asymmetric profile, with one horn almost completely missing, so the symmetric model is not able to match the data.

By visual inspection we identify two galaxies out of the sample for which the RC and flux profile are obviously inconsistent under the basic assumptions of our model, with the fit completely failing to match the basic width of the flux profile. F568-V1 has large wings in the flux profile at large velocity, which may be due to merging gas which our model cannot account for. Despite having a quality flag of 2, F565-V2 appears to have an RC with too high a velocity given the narrowness of the observed flux profile. However, due to the large uncertainties, both galaxies are a reasonable fit in terms of MSE (although F568-V1 requires $\sighi\approx100$ km/s to match the wings) due to the suspiciously large $\sigma_\text{rms}$. We show both galaxies in Appendix \ref{sec:excluded}. 

In Table \ref{tab:results}, we show basic properties, the MSE statistic and constrained parameter values for all galaxies in the sample.

\begin{table*}
\caption{Basic galaxy properties, including the \HI{} mass measured by ALFALFA survey and tabulated in the SPARC database respectively, the goodness-of-fit statistic MSE (equation \ref{eq:residual}), and the inferred values of $\hiflux$ and $\sighi$ for the sample. Each \HI{} mass has a 10\% calibration uncertainty.}
\centering
\begin{tabular}{ccccccll}
\hline
Name & \#AGC & SPARC log($M_*/M_{\odot}$)  & SPARC log($M_{HI}/ M_{\odot}$) & ALFA log($M_{HI}/ M_{\odot}$) & MSE & $\hiflux$ (Jy km/s) & $\sighi$ (km/s)\\
\hline
DDO154 & 8024 & 7.42 & 8.44 & 8.5 & 57.7 & 82.0 ± 0.005 & 11.0 ± 0.0031 \\
F563-V2 & 180739 & 9.17 & 9.34 & 9.42 & 0.171 & 3.2 ± 0.0061 & 13.0 ± 0.093 \\
F565-V2 & 196124 & 8.45 & 8.84 & 8.92 & 1.55 & 1.5 ± 0.0039 & 24.0 ± 0.19 \\
F568-V1 & 201046 & 9.28 & 9.4 & 9.71 & 2.98 & 5.1 ± 0.00017 & 96.0 ± 0.18 \\
F571-V1 & 217484 & 8.97 & 9.09 & 9.22 & 0.343 & 1.2 ± 0.0038 & 14.0 ± 0.11 \\
F574-1 & 220862 & 9.51 & 9.55 & 9.65 & 0.625 & 2.0 ± 0.0056 & 17.0 ± 0.18 \\
F579-V1 & 240454 & 9.77 & 9.35 & 9.46 & 0.299 & 1.4 ± 0.0034 & 7.0 ± 0.099 \\
IC4202 & 8220 & 11.0 & 10.1 & 10.0 & 0.865 & 4.6 ± 0.0068 & 26.0 ± 0.12 \\
NGC4559 & 7766 & 9.99 & 9.76 & 9.75 & 181.0 & 300.0 ± 0.0069 & 14.0 ± 0.0015 \\
NGC7814 & 8 & 10.6 & 9.03 & 8.97 & 2.4 & 18.0 ± 0.0079 & 16.0 ± 0.029 \\
UGC00634 & 634 & 9.17 & 9.56 & 9.55 & 5.63 & 16.0 ± 0.004 & 11.0 ± 0.019 \\
UGC05005 & 5005 & 9.31 & 9.49 & 9.7 & 2.54 & 7.9 ± 0.0043 & 14.0 ± 0.042 \\
UGC05721 & 5721 & 8.42 & 8.75 & 8.67 & 94.9 & 52.0 ± 0.0057 & 14.0 ± 0.0047 \\
UGC05750 & 5750 & 9.22 & 9.04 & 9.39 & 1.02 & 3.2 ± 0.0043 & 14.0 ± 0.078 \\
UGC05999 & 5999 & 9.23 & 9.31 & 9.61 & 0.924 & 7.5 ± 0.0042 & 12.0 ± 0.026 \\
UGC06786 & 6786 & 10.6 & 9.7 & 9.59 & 3.14 & 18.0 ± 0.0072 & 12.0 ± 0.026 \\
UGC07261 & 7261 & 8.94 & 9.14 & 9.06 & 5.99 & 29.0 ± 0.0043 & 13.0 ± 0.0055 \\
UGC11820 & 11820 & 8.69 & 9.3 & 9.29 & 19.2 & 24.0 ± 0.0042 & 11.0 ± 0.0095 \\
UGC11914 & 11914 & 10.9 & 8.95 & 8.81 & 1.11 & 9.3 ± 0.0047 & 6.5 ± 0.028 \\
UGC12732 & 12732 & 8.92 & 9.56 & 9.54 & 84.2 & 83.0 ± 0.0048 & 9.5 ± 0.0031 \\
\hline
\end{tabular}\label{tab:results}
\end{table*}

\subsection{Sample analysis}\label{sec:sample}

\begin{figure*}
    \includegraphics[width=\textwidth]{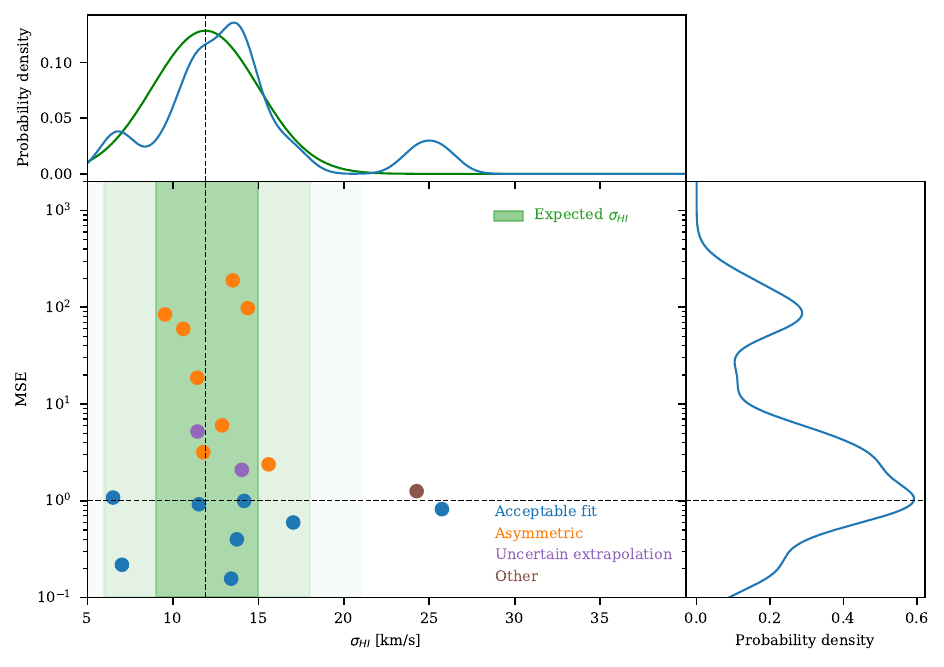}
    \caption{A plot of the goodness-of-fit statistic MSE (equation \ref{eq:residual}) against the preferred $\sighi$ from our model fit, as well as their respective 1D kernel density estimate plots (blue lines, smoothed by Gaussian 1 km/s for $\sighi$ and 0.2 dex for MSE for clarity). The green contours and line correspond to expectations from literature of $\sighi = 12 \pm 3$ km/s from literature \citep{mogotsiHICOVelocity2016}. The uncertainties on $\sighi$ are very small, so are not shown. Given the approximations of the model, most galaxies are a decent fit. The distribution of $\sighi$ peaks as expected (no prior applied is applied in the inference), but there are a couple of galaxies with higher $\sighi$, suggesting either the gas dispersion in these galaxies is higher than expected and/or the fit is compensating for additional effects that broaden the flux profiles. We visually inspect the significant outliers, which we colour according to the most probably cause: an asymmetric flux profile (orange) or uncertain RC or $\Sighi$ extrapolation (purple), compared to other surveys targeting the same galaxy. The classification is described fully in Section \ref{sec:sample}.  Of the two unusual galaxies described in Section \ref{sec:overview}, F565-V2 is labelled "Other" and F568-V1 (which has $\sighi \approx 100$ km/s) is not shown. The black dashed lines indicate the centre of the $\sighi$ prior and the reference value MSE=1. \label{fig:dbar_empiricall}}
\end{figure*}

We quantify the consistency of the model flux profile with observations for the whole sample using the MSE statistic (equation \ref{eq:residual}). In Fig.\ref{fig:dbar_empiricall} we plot the goodness-of-fit statistic MSE against the preferred $\sighi$ from the fit to the flux profile. Whilst the peak $\sighi$ is consistent with expectations, there are a couple of galaxies with higher $\sighi$ which is not expected, which on the face of it suggests either the gas dispersion in these galaxies is higher than expected, or possibly that there are additional sources of noise that broaden the profile or it's compensating for the model being a bad fit to the data in any other way. Whilst for many galaxies the difference between the data and model is reasonable given the limitations of the data and the simplifications of the model, for some galaxies there is a large deviation. Often this is driven by the very small statistical uncertainties on the flux profile, which are a consequence of the high signal-to-noise ratio (SNR, which has an extremely strong correlation with MSE). This suggests that, for our fits, as the SNR increases systematic errors are amplified. We find no significant correlation between MSE and $\sighi$.

\begin{figure*}
    \includegraphics[width=\textwidth]{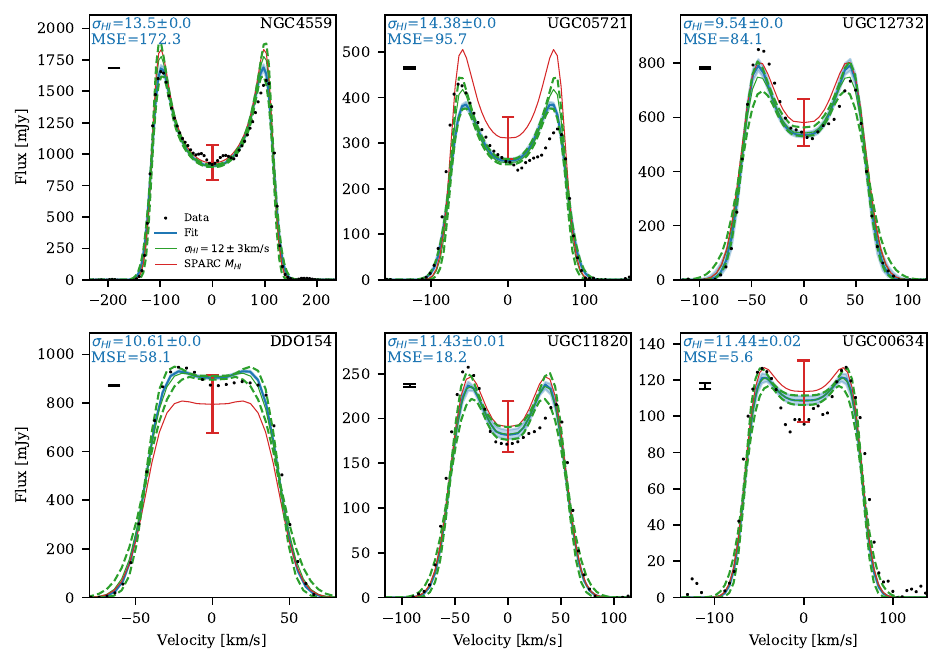}
    \caption{The same as the middle panels of Fig. \ref{fig:empirical-fit}, but for the galaxies for which the model is the a poor match to data according to the mean squared error statistic (which is labelled in blue). From top left are the five worst fits, which are all due to asymmetry in the flux profile (as described in Section \ref{sec:sample}). UGC00634 is shown as an example of a galaxy with a uncertain extrapolation of $\Sighi$.}
    \label{fig:badfits}
\end{figure*}

We investigate cases of inconsistency between the model flux profile model and the data by visual inspection. We identify two primary causes for large values of MSE:

\begin{enumerate}
    \item \textbf{Uncertain extrapolation:} For some galaxies the measured RC in SPARC does not extend as far as the \HI{} flux, so the flux profile is sensitive to the uncertain extrapolation of the RC.
    \item \textbf{Asymmetric:} for some galaxies the flux profile is strongly asymmetric, for example missing one horn. We quantify asymmetry using the $\Delta \mathrm{BIC}_{\mathrm{asym}}$ metric (equation \ref{eq:bic}). We label galaxies as asymmetric rather than missing flux if we cannot find a significantly discordant observation in the EDD.
\end{enumerate}
The above classifications are shown as colours in Fig. \ref{fig:dbar_empiricall}. We show all galaxies with $\Delta \mathrm{BIC}_{\mathrm{asym}} > 10$. We only inspect galaxies with high MSE (>1.5) or high $\sighi$ (>20 km/s) for uncertain extrapolation (at lower SNR identifying this pathology become ambiguous).

In Fig.~\ref{fig:badfits} we show the five worst fits defined by MSE (which are all asymmetric), as well as UGC00634, which has uncertain extrapolation. We see NGC4559 has a shape that is well matched by the model, but due to the tiny uncertainties the MSE is very large and the profile is statistically asymmetric. UGC-5721, UGC12732, DDO154 and UGC11820 are all strongly asymmetric by our criteria, due to missing horns. \citeauthor{hoffmanTotalALFALFANeutral2019} identify DDO154 as a galaxy for which the reanalysis pipeline still appears to be missing some flux (although less than the original pipeline) when compared to large-beam telescope observations from literature. This suggests that some asymmetries may still be artificial in the reanalysed spectra.

\begin{figure*}
    \includegraphics[width=\textwidth]{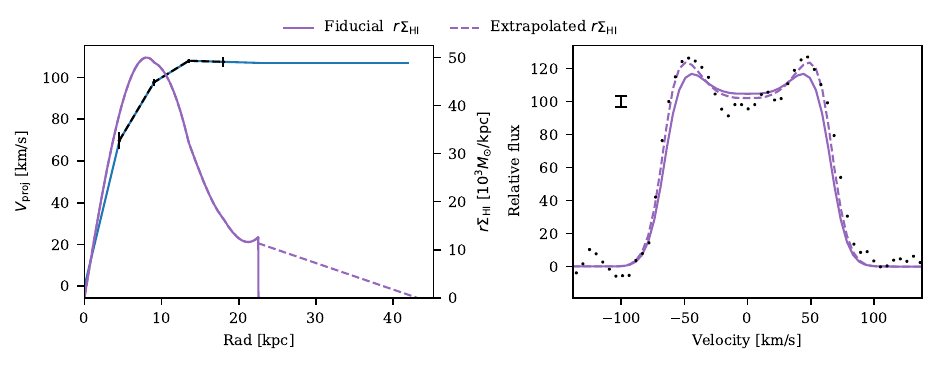}
    \caption{The same as the left and middle panels of Fig. \ref{fig:empirical-fit}, but for UGC00634 only, a galaxy we identified as possibly having a poor fit due to the uncertain extrapolation of the \HI{} surface density profile. For the fiducial unextrapolated $r \Sighi$, the best fit model (purple, uncertainties not shown) is a poor match to the data. Extrapolating $r \Sighi$ provides a better match to the data as it sharpens the horns of the profile. We test similar extrapolation for all galaxies, but find for most galaxies the profile is similar to no extrapolation.}
    \label{fig:extrapolation}
\end{figure*}

We chose not to extrapolate $\Sighi$ in our model due to the uncertainty, although $r\Sighi$ often appears to be declining linearly towards zero (see Fig.~\ref{fig:empirical-fit}). Comparing the total flux when doing a linear extrapolation of $r$$\Sighi$ and when doing no extrapolation, we find the difference is less than 2.5\% of the total flux for all but two galaxies (for which it is $\sim5\%$ ). We also see the shapes of the resulting flux profiles are similar. Therefore this potential bias is likely not significant. We identify UGC00634 as a galaxy that may have its fit quality significantly affected by the uncertain extrapolation Fig.~\ref{fig:extrapolation}.

\begin{figure}
    \includegraphics[width=\columnwidth]{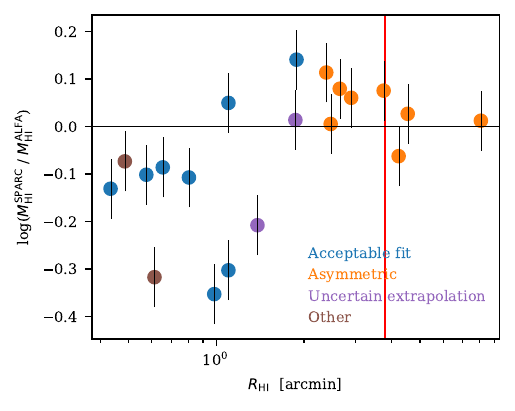}
    \caption{A plot of the ratio of the SPARC and ALFALFA \HI{} mass measurements against the size of the galaxy's \HI{} disc (defined as the radius at which the surface density drops drops to 1 $\mathrm{M}_{\odot} / \mathrm{pc}^2$). The red line shows the approximate ALFALFA beam width (which is roughly elliptical with a half power beam width of 3.3 x 3.8 arcmin). The galaxies are coloured by the same classification as in Fig.~\ref{fig:dbar_empiricall}. For large angular size galaxies, the measurements are fairly consistent, with some evidence for missing flux from the ALFALFA profile. At low angular size the SPARC measurement misses some flux due to the low surface density of gas in the outer parts of these low mass galaxies.}
    \label{fig:compare_mass}
\end{figure}

In Fig.~\ref{fig:compare_mass} we compare the SPARC and ALFALFA \HI{} mass measurements, assuming the SPARC distance for both. For galaxies with large \HI{} discs, which tend to be larger spiral galaxies, the two measurements are fairly similar, but with some evidence for missing flux from the ALFALFA spectrum even with the upgraded pipeline.  Therefore some of the asymmetries in the ALFALFA spectra may be a measurement artifact rather than intrinsic. For some galaxies that appear smaller on the sky (mostly dwarf galaxies), the SPARC analysis misses flux, likely due to the low surface brightness in their outskirts. However there is still statistical agreement between RC and flux profile, as these fainter galaxies have larger uncertainties on both RC and flux profile measurements. Our linear extrapolation of $r \Sighi$ is not able to resolve the discrepancy between SPARC and ALFALFA masses for these galaxies. This suggests the $\Sighi$ profiles may have a more complex shape beyond the final measured radius in SPARC that we were not able to capture with a simple extrapolation model.

\begin{figure}
    \includegraphics[width=\columnwidth]{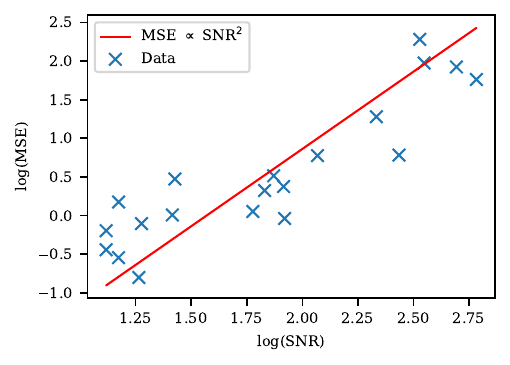}
    \caption{A plot of log(SNR) against log(MSE) for the sample, with the best fit line with MSE $\propto$ $1/\sigma_\text{rms}^2$ $\propto$ SNR$^2$ shown in red. The scaling is roughly consistent with high MSE being driven by poor fits due to features such as asymmetries that are amplified by high SNR.}
    \label{fig:snr_vs_mse}
\end{figure}

Ideally one would study the dependence of MSE with properties such as asymmetry, conditioned on SNR. However, likely due to the small sample size and the varied factors driving poor fits, no significant secondary correlations could be found. In Fig. \ref{fig:snr_vs_mse} we show the scaling of MSE with SNR for the sample. If MSE increased only due to the reduction in statistical uncertainty as SNR increases, then we would expect a MSE $\propto$ SNR$^2$ scaling (as SNR $\propto$ $1/\sigma_\text{rms}$), which is approximately consistent with the data. This indicates that higher-SNR observations do not typically have a lower absolute discrepancy between RC and flux profile.

\section{Discussion}\label{sec:discussion}

\subsection{Interpretation of results}

We have used a simple model (Eq.~\ref{eq:convolution}) that combines the azimuthally-averaged SPARC RC and \HI{} surface density profiles under the assumption of axisymmetry to assess the consistency of \HI{} flux profiles and resolved rotation curves. We find that for most galaxies the model flux profile is an acceptable fit to the data. Some deviation is a priori expected due to limitations of the data:
\begin{itemize}
    \item The observed RC profile does not extend as far as the observed $\Sighi{}$ profile, necessitating extrapolation.
    \item The integrated flux profile may pick up flux from beyond the extent of the measured $\Sighi{}$ profile.
    \item RC measurements are known to be imperfect, with uncertainties that may be under- or over-estimated \citep[e.g.][]{sellwoodUncertaintiesGalaxyRotation2021}.
\end{itemize}

However, for some galaxies we found significant deviation between model and data. In six cases we identify this as being due to asymmetric flux profiles. The SPARC galaxies are mostly nearby, and thus have high signal-to-noise in ALFALFA.

This amplifies systematics, so even modest asymmetries could lead to large statistical discrepancies. Flux profiles are usually studied in lower signal-to-noise regimes, such as smaller or more distant galaxies, where one might expect the statistical uncertainty to become dominant. On the other hand physical asymmetries are likely less prevalent in SPARC than the ALFALFA sample, as the SPARC galaxies are selected to be regular to enable detailed kinematic studies.

Even in some galaxies with an acceptable model fit, we found cases where a slightly higher gas dispersion than expected was preferred. This may be the result of additional broadening of the flux profile due to physical processes, such as warps. The horns are coherent features, so it is not surprising that any deviation from a flat RC would dampen them. One plausible reason is that small scale wiggles in the RC (e.g. caused by spiral arms) remove the coherence of the horn. This would not be captured by an azimuthally-averaged RC. However the velocity uncertainties, which are dominated by the difference between approaching and receding sides of the RC for most galaxies, are fairly small. In fact for many galaxies the uncertainty in the line-of-sight velocity is less than the velocity dispersion across much of the RC.

Pan et. al (in prep.) compare the amount of \HI{} flux measured by the interferometric MIGHTEE-HI survey \citep{ponomarevaMIGHTEEHBaryonicTully2021} and FAST. They find on average the FAST galaxies are $7\pm 2 \%$ more massive, which they attribute to \HI{} in the circumgalactic medium or extragalactic gas. This relatively close agreement suggests that current and future pipelines designed for resolved observations can obtain flux profiles consistent with unresolved single dish studies. This suggests the problem of missed flux for resolved sources found in the original ALFALFA pipeline will not be an obstacle to flux profile analyses in future surveys. We also found no obvious evidence for extragalactic gas, high velocity clouds or satellites in the SPARC flux profiles, other than possibly the wings of F568-V1 (see Appendix \ref{sec:excluded}).

\subsection{Future prospects}

Current and future surveys will obtain vast numbers of flux profiles in regimes inaccessible to \HI{} rotation curve studies, including at high-redshift and for cosmological volumes of dwarf galaxies for the first time. Our work reveals how samples of flux profiles can be selected as suitable for axisymmetric modelling, enabling the maximum amount of information on their dynamics to be extracted. This is in contrast to current analyses that use highly compressed linewidth summary statistics. Harnessing the full flux profile will enable studies of the galaxy-halo connection to be extended into unprecedented regimes, and new tests of gravity more broadly. 

There are several ways this study could be improved with future data sets. The main one is to use larger galaxy samples to gain a statistical understanding of what type of galaxy (or observation) produces flux profiles and RCs that are consistent with each other. The arrival of new surveys such as FAST \citep{zhangFASTAllSky2024}, WALLABY \citep{koribalskiWALLABYSKAPathfinder2020} and MIGHTEE \citep{maddoxMIGHTEEHIHIEmission2021} will provide a wealth of new data on the \HI{} content of galaxies, including large numbers of flux profiles.

A more sophisticated treatment could be carried out using a separate $\Sighi$ and RC for the approaching and receding side of the disk, data which are not available for SPARC. This would allow us to ascertain to what extent asymmetries in the \HI{} distribution and RC manifest as significant asymmetries in the flux profile compared to observational effects such as the one we have highlighted. Similarly, analysing resolved radio datacubes, cutting and binning them in such a way as to mimic the integrated observations, would allow for a more detailed study of missing flux.

Studying a survey such as MIGHTEE, which produces RCs for more distant galaxies, would allow a better comparison of flux profiles and RCs for galaxies at distances more commonly probed in unresolved studies. It would also be interesting to study a lower mass sample such as Little THINGS \citep{hunterLittleThings2012,iorioLITTLETHINGS3D2016} in which the galaxies have higher dispersion.

Finally, it would be interesting to apply our method to the integrated flux profiles of other tracers. For example, a similar flux profile model was previously used to study the Tully--Fisher relation using fibre H$\alpha$ profiles \citep{moczTullyFisherRelation2012}. This could be particularly impactful on studies of high-redshift galaxies (for example using submillimetre observations, \citealt{jonesALPINEALMAIiSurvey2021}), where often resolved observations are not possible, and the flux profile is the only available information.

\section{Conclusions}\label{sec:conclusions}

We lay the groundwork for integrated \HI{} flux profiles to be used as probes of dark matter distributions and modifications to standard dynamics in galaxies. This is achieved by combining the azimuthally-averaged RC and \HI{} surface density profiles of galaxies in the SPARC sample to generate model flux profiles which we then compare to the ALFALFA data. We develop an objective Bayesian criterion to measure galaxy asymmetry while accounting for flux profile uncertainties, enabling a robust determination of whether our axisymmetric modelling suffices. Our main findings are:

\begin{itemize}
\item{Within the limitations of the data, set by the limited extent of the rotation curve and surface density measurements, the \HI{} rotation curve and the flux profile are in good statistical agreement for most of the sample. For some galaxies strong asymmetries prevent our symmetric model from matching the data.}
\item{With $\sighi$ as a free parameter in our model with a uniform prior, we found that the preferred value is in good agreement across the sample with literature expectations from resolved studies, with a peak at $14$ km/s.}
\item{Our asymmetry statistic can be applied to galaxies for which RC data is not available to determine their suitability for dynamical or other types of modelling based on the assumption of axisymmetry.}
\end{itemize}
We therefore identify conditions under which the \HI{} flux profile is not a good tracer of the RC, and show for the rest that the two tracers have high statistical consistency. The cheaper and more accessible flux profiles may therefore be used for detailed physical tests based on galaxy dynamics.

\section*{Acknowledgements}

We thank Federico Lelli, Anastasia Ponomareva, Hengxing Pan and Matt Jarvis for useful inputs and discussion. We thank Martha Haynes and Lyle Hoffman for sharing the ALFALFA data, and Federico Lelli the \HI{} surface densities of the SPARC galaxies. We also thank Chuan-Peng Zhang for sharing their FAST data.

TY acknowledges support from a UKRI Frontiers Research Grant [EP/X026639/1],
which was selected by the ERC. HD is supported by a Royal Society University Research Fellowship (grant no. 211046). This project has received funding from the European Research Council (ERC) under the European Union's Horizon 2020 research and innovation programme (grant agreement No 693024).

For the purpose of open access, the authors have applied a Creative Commons Attribution (CC BY) licence to any Author Accepted Manuscript version arising.

\section*{Data Availability}

Data from the ALFALFA survey is located at \url{http://egg.astro.cornell.edu/alfalfa/data/index.php}. The SPARC database is located at \url{http://astroweb.cwru.edu/SPARC/}. Other data will be made available on reasonable request to the corresponding author.



\bibliographystyle{mnras}
\bibliography{paper} 




\appendix

\section{Unusual galaxies}\label{sec:excluded}

\begin{figure*}
    \includegraphics[width=0.97\textwidth,height=0.18\textheight]{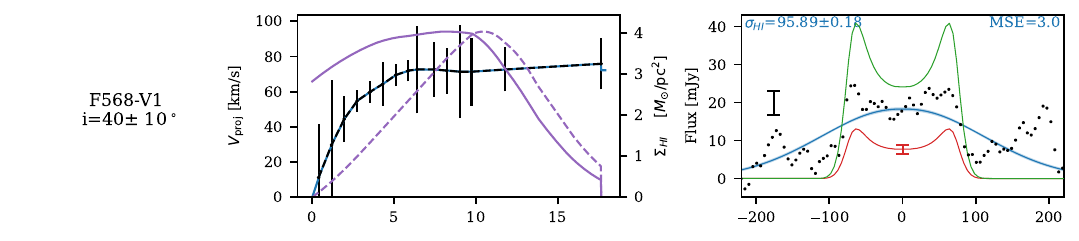}
    \includegraphics[width=0.97\textwidth,height=0.18\textheight]{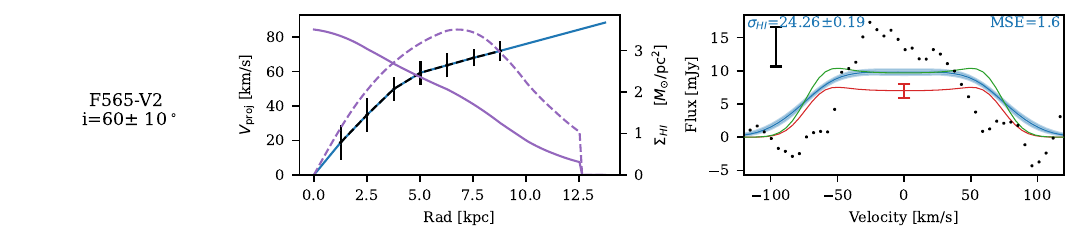}
    \caption{The same as Fig.~\ref{fig:empirical-fit} but for the two unusual galaxies for which our model cannot even match the basic width of the flux profile. For F568-V1 the flux profile has wing features that cannot be explained in our model, possibly due to merging gas. For F565-V2 the RC normalisation is too high to generate a flux profile of the observed narrow width.}
    \label{fig:excluded}
\end{figure*}

In Fig.~\ref{fig:excluded} we show the two unusal galaxies described in Section \ref{sec:overview}.

\bsp	
\label{lastpage}
\end{document}